\PassOptionsToPackage{svgnames}{xcolor}
\documentclass[]{article}

\usepackage{iclr2018_conference}
\iclrfinalcopy

\usepackage[utf8]{inputenc} 
\usepackage[T1]{fontenc}    
\usepackage[]{hyperref} 
\AtBeginDocument{
  \hypersetup{
    citebordercolor=Violet,
    linkbordercolor=Cyan,   
    urlbordercolor=Magenta}}
\usepackage{url}            
\usepackage{booktabs}       
\usepackage{amsfonts,amsmath}       
\usepackage{nicefrac}       
\usepackage{microtype}

\usepackage[backend=bibtex, 
            style=numeric-comp,
            citestyle=numeric-comp,
            sorting=none,
            doi=false,
            url=false,
            giveninits=true,
            maxbibnames=47]{biblatex}

\bibliography{references}

\usepackage{aliases}
\usepackage{misc}
\usepackage{jhh-misc2-nips}
\title{\centering
Reconstructing probabilistic trees \\ of cellular differentiation \\ from
single-cell RNA-seq data
}

\author{
  Miriam Shiffman\textsuperscript{ 1,2,}
  \thanks{contact: \texttt{shiffman@\{mit.edu,broadinstitute.org\}}} \! ,
  William T. Stephenson\textsuperscript{ 1},
  Geoffrey Schiebinger\textsuperscript{ 1,2},
  \\[.05cm]
  \textbf{
  Jonathan Huggins\textsuperscript{ 3},
  Trevor Campbell\textsuperscript{ 4},
  Aviv Regev\textsuperscript{ 1,2,5},
  Tamara Broderick\textsuperscript{ 1}
  }
  \\[.2cm]
  \textsuperscript{1 }MIT
\\
 \textsuperscript{2 }Klarman Cell Observatory, Broad Institute of MIT \& Harvard
\\
\textsuperscript{3 }Department of Biostatistics, Harvard
\\
\textsuperscript{4 }University of British Columbia 
\\
\textsuperscript{5 }Howard Hughes Medical Institute
}

\begin{document}

\maketitle
\begin{abstract}

Until recently, transcriptomics was limited to bulk RNA sequencing,
obscuring the underlying expression patterns of individual cells in favor
of a global average. Thanks to technological advances, we can now profile
gene expression across thousands or millions of individual cells in
parallel.
This new type of data has led to the intriguing discovery that individual
cell profiles can reflect the imprint of time or dynamic processes.
However, synthesizing this information to reconstruct dynamic biological
phenomena from data that are noisy, heterogenous, and sparse---and from
processes that may unfold asynchronously---poses a complex computational
and statistical challenge.
Here, we develop a full generative model for probabilistically
reconstructing trees of cellular differentiation from single-cell RNA-seq
data.
Specifically, we extend the framework of the classical
Dirichlet diffusion tree to simultaneously infer branch topology and latent
cell states along continuous trajectories over the full tree. 
In tandem, we construct a novel Markov chain Monte Carlo sampler that 
interleaves Metropolis-Hastings and message passing to leverage 
model structure for efficient inference.
Finally, we demonstrate that these techniques can recover latent trajectories
from simulated
single-cell transcriptomes.
While this work is motivated by cellular differentiation, 
we derive a tractable model that provides
flexible densities for any data 
(coupled with an appropriate noise model)
that arise from continuous
evolution along a latent nonparametric tree.

\end{abstract}

\section{Introduction}

Many fundamental questions in biology invoke the question of how to describe and
measure cell ``state''---that is, the differences in identity between two cells
with the same genome. One particularly informative measure is \emph{gene
expression}, i.e.\ the amount of RNA transcripts per gene.
Recent techniques
offer unprecedented biological insight by enabling massively-parallel
quantification of RNA molecules at single-cell resolution (\emph{single-cell RNA
sequencing}, or scRNA-seq)~\cite{DropSeq, droplet, 10X}. However, the resulting
data are noisy and zero-inflated, confounding traditional
analyses~\cite{sc-rev-regev, sc-rev-teichmann}. In this work, we employ a
Bayesian approach to directly model sources of uncertainty in single-cell
transcriptomic data and infer interpretable, probabilistic insight into cell state.

The particular biological phenomenon we study is \emph{cellular
differentiation}, the process by which a less specialized progenitor
(e.g., a stem cell) gradually gives rise to cells with more specialized
function.
More concretely, the initial cell division is generally asymmetric,
leaving one daughter cell with its stemness intact while the other cell
goes on to seed a lineage of increasingly committed cells.
This dynamic process can be represented as a tree, whose branches
designate the incremental progression of more potent cells into various
mature cell types~\cite{devbio}.
While the lineage of a single cell can be traced back through a path of
binary cell divisions, the tree we describe reflects the changing latent
\emph{potential} of a cell to give rise to 
more differentiated cells and cell fates,
a dynamic process that inherently dwells along a spectrum.
The importance of understanding this phenomenon is underscored by its
ubiquity---for example, every cell in the human body arose through
differentiation and many are continually replenished through this process, on
the order of days to years~\cite{devbio,development,hematopoiesis,gut}.
Yet, fundamental questions remain.
How do identical progenitors reliably give rise to a suite of diverse 
cell fates? How do the dynamics of gene expression change over time and
across lineages? What is the molecular program by which orchestrated
changes in expression lead cells down one path over another---and how
deterministic is this decision?

These systems-level questions are driven and controlled by molecular
mechanisms involving coordinated regulation of gene expression.
The levers of this 
coordination are transcription factors,
whose combinatorial interactions together compose higher-level
transcriptional ``programs''
that enact changes in cellular identity and state~\cite{devbio,sc-rev-regev}.
Ultimately, we seek to gain insight
into this lower-dimensional subspace (of molecular circuitry) that
generates observed expression profiles---that is, how interactions among
a handful of transcription factors orchestrate complex patterns of
expression
to drive consistent trajectories of differentiation.

To this end, we seek to infer the latent tree of cellular differentiation, and
the genes that drive its topology, from scRNA-seq measurements (noisy snapshots
of cell state). Since this assay is destructive to cells we cannot
follow a single cell along its trajectory through time, and must instead infer
this dynamic process by sampling many static snapshots of individual
cells~\cite{sc-rev-regev, sc-rev-teichmann, sc-rev-trajectory}. Further
complicating analysis, any given sample of cells represents a non-uniform draw
of unlabeled time points from the underlying tree of differentiation, which
unfolds asynchronously~\cite{sc-rev-trajectory}.

In this work, we begin by reviewing previous approaches
to cell trajectory
reconstruction and Bayesian inference on trees,
including their potential limitations,
in \Cref{sec:background},
In \Cref{sec:model}, we develop a novel generative model for scRNA-seq
count data arising from cells undergoing a dynamic, bifurcating differentiation
process. In particular, we extend the existing \emph{Dirichlet diffusion tree} model~\cite{ddt,ddt2} to
incorporate a continuous distribution over pseudotime from root to leaves,
filling a hole in Bayesian nonparametrics for probabilistic trees where data are
not confined to discrete nodes~\cite{ddt,ddt2,pydt,coalescent,hdp,nCRP,tssb}.
In \Cref{sec:inference}, we describe our novel MCMC algorithm, which
interleaves Metropolis-Hastings and message passing steps to leverage model structure
(and variable augmentation) for efficient inference.
Finally, in \Cref{sec:results}, we present initial experiments that
demonstrate the ability of our techniques to recover structure from
simulated
single-cell transcriptomes.

\section{Background} \label{sec:background}

\subsection{Single-cell trajectory reconstruction}

Many methods exist to infer lineage relationships from single-cell
expression profiles~\cite{sc-rev-trajectory}.
However, most retain no notion of uncertainty,
do not infer differentially expressed genes in conjunction with
reconstructing the tree,
or require suitable normalization or dimensionality
reduction beforehand~\cite{dpt,wishbone,monocle,monocle2,phate,bgp,slicer,scuba,mfa,gpfates,sc-compare}.
Moreover, the latter preprocessing decisions are generally \emph{ad hoc}
rather than reflecting an explicit model of data collection---compounding
the lack of a framework for uncertainty. 
Many methods also assume a fixed number of branch
points (which must be specified)~\cite{monocle,wishbone,gpfates,mfa,sc-compare},
while others assume perfect time collection information and the ability
to precisely infer cell growth~\cite{ot}.
Finally, adding new data from the same system is generally nontrivial,
and often requires recomputation from scratch~\cite{sc-compare}.

In contrast, we develop a Bayesian model of differentiation
that is not fragile to preprocessing choices,
since we directly model gene expression counts, and
yields interpretable results for differential expression across lineages.
Our approach provides a generative means of evaluating
and simulating from the model, as well as a principled way to account for 
technical factors like zero-inflation due to gene dropout.
The modularity of the generative model facilitates (streaming)
incorporation of heterogenous datasets (e.g.\ samples enriched for stem
cells versus mature cells),
the ability to leverage cell time information as available,
and extensions to multiple modalities of
observation (e.g.\ multi-omic data beyond scRNA-seq).
Further, we leverage Bayesian nonparametrics 
to learn a flexible model that expands as needed
with expanded sequencing data, permitting trees of unbounded width and depth
with no requirement for \textit{a priori} knowledge of the number of cell fates.

\subsection{Bayesian inference} \label{sec:bayes}

Bayesian inference relies on writing out a full generative model---the
prior $p(\theta)$, encoding our beliefs about the
uncertainty structure over
parameters $\theta$, and the likelihood $p(x \mid \theta)$, which couples
the observations $x$ to the model through their dependence on the latent
settings of $\theta$. Then, inference uses Bayes' rule to invert the
generative model into a mechanism for sampling from its posterior
$p(\theta \mid x)$ (up to a normalizing constant, the model evidence
$p(x)$, which is generally intractable). The posterior represents the
density over the hidden parameters of interest, given the observations we
have in hand. Concretely, here we observe cell expression profiles ($x$),
and aim to recover the topology, times, and latent states of all cells
and their underlying tree of differentiation potential (collectively,
$\theta$).

\subsection{Probabilistic latent tree models}

Previous approaches to Bayesian modeling of latent tree densities
assume data are generated either only at the leaves
(e.g.\ Dirichlet~\cite{ddt,ddt2} or Pitman--Yor~\cite{pydt} diffusion trees,
hierarchical Dirichlet process~\cite{hdp}, Kingman's
coalescent~\cite{coalescent})
or only at the nodes of the tree (nested Chinese restaurant
process~\cite{nCRP}, tree-structured stick breaking process~\cite{tssb})
(\Cref{app-figs}, \Cref{fig-3trees}, left and center). In contrast, we
seek to model data arising from a more challenging regime in which
observations are generated continuously over the entire tree, from root
to leaves (\Cref{app-figs}, \Cref{fig-3trees}, right).
Further, we seek the flexibility to model data points 
not as uniformly populating the tree, but rather as sporadic or
dense in relation to both the ``speed'' at which cells undergo each
transition and the relative number of cells that follow each path.

\newcommand{\figddt}{
\begin{figure} 
  \centering
  \vspace{-5pt}
  \includegraphics[width=.6\textwidth]{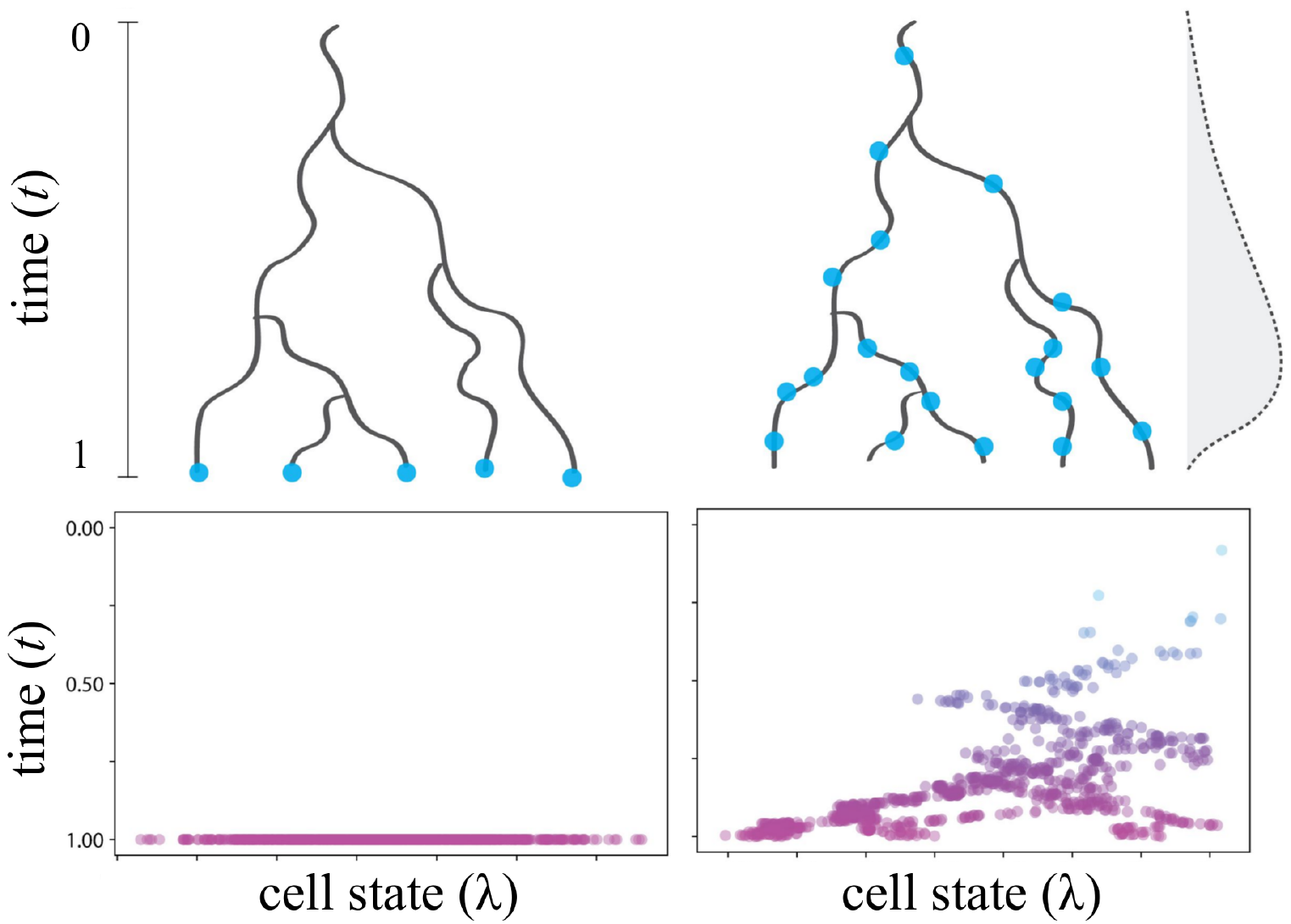}
    \caption{
        Classical (\emph{left}) vs.\ augmented DDT (\emph{right}).
        \emph{Upper}: cartoon of tree and data for 1-d cell state
        (\emph{middle left \& middle right}), with
        $\distBeta(5, 1)$ prior over cell pseudotimes in our augmented DDT
        (\emph{far right}).
        \emph{Lower}: Latent cell states simulated from each model (colored by time).}
    \label{fig-ddt}
\end{figure}
}

\section{Bayesian generative model for cellular differentiation} \label{sec:model}

Here, we describe how a cell's observed profile in gene expression space
(transcript counts $x_\icell$) is generated based on its latent cellular
differentiation state ($\lambda_\icell$), where $\icell$ indexes each
cell in our single-cell RNA-seq dataset.
In \Cref{sec:ddt}, we review a previous model, the Dirichlet diffusion
tree~\cite{ddt,ddt2}, that provides flexible tree-structured densities. This model serves as a springboard for depicting 
cell trajectories in terms of latent binary branches.
Then in \Cref{sec:ourddt}, we create a new model capable of representing
latent cell states as draws from continuous trajectories over a
probabilistic tree. Transformations of these latent states
$\lambda_\icell$ then directly parameterize the observation model over
gene expression profiles $x_\icell$ from differentiating cells.
Finally, in \Cref{sec:obs_model}, we describe this model for generating
noisily observed transcript counts from their ``true'' underlying values
in each cell.

\subsection{Dirichlet diffusion trees} \label{sec:ddt}
Dirichlet diffusion trees (DDTs) provide a family of nonparametric priors over
exchangeable data that arise from a latent branching process. The generative
model yields tree topologies and branch times (via a hazard process) as well as
latent states at and along branches (via Gaussian diffusion)~\cite{ddt,ddt2}
(\Cref{app-ddt}).
While simple in composition,
the DDT has the capacity to model flexible densities; 
for example,
its density estimation properties were previously shown to outperform
both a Dirichlet process mixture model and
Gaussian process density sampler~\cite{gpds}.

Specifically, branch times and topologies are generated by iteratively
simulating particles that follow a self-reinforcement scheme and deviate
from existing paths according to a probabilistic divergence function, or
branching rate (\Cref{app-ddt}). This function defines a hazard process
(such that every leaf must branch by unit time) 
parameterized by a single
scalar $\conc$: a (positive) smoothness parameter governing whether branches are
concentrated toward the root or the leaves. Following Neal~\cite{ddt}, we
assume a branching rate of $\branchrate(t) = \conc / (1-t)$. This function
defines the instantaneous chance of divergence, 
$\branchrate(t)\, \d t / \nparticles$,
where $\nparticles$ is the number of particles that have previously traversed
the given branch without diverging.
Then, let 
\[
  \cbranchrate(t) \defined \int_0^t \branchrate(u) \d u = -\conc\log(1-t);
\]
this is the cumulative branching function. If a particle is on an
existing leg of the tree bookended by times $[t_a, t_b]$ and
$\nparticles$ particles have previously traversed this path, the
likelihood of branching by some time $t > t_a$ is defined by a Poisson
process:
\[
  \pbranchonleg_{t_a}(t) \defined \Pr(\mathrm{branch \ in } \ [t_a, t])
  = 1 - e^{-(\cbranchrate(t)-\cbranchrate(t_a)) / \nparticles}
  = 1 - \left( \frac{1-t}{1-t_a} \right)^{\conc / \nparticles}.
\]
To determine if/when a new particle diverges along this branch, we calculate
$t'$
by the inverse CDF method. If
$t'$ is past the end of the branch, $t_b$,
then the particle
does not yet diverge and instead follows one of the existing child branches at
time $t_b$ (where branch choice proportionally favors the child that previous
particles have chosen).
Following Knowles \emph{et al.}~\cite{varddt}, we calculate efficient likelihoods for
this exchangeable distribution over topologies 
by reformulating the likelihood
as a simple function of (cached) harmonic numbers.

Node locations (as well as the branches themselves) can then be sampled
according to a Brownian motion process.
Specifically, a particle that has reached $X(t)$ at time $t \in (0, 1)$
will diffuse to $X(t + \d t) = X(t) + \distNorm(0, \; \DDTvar \, \I \cdot
\d t)$ after an infinitesimal amount of time $\d t$, for some base
variance $\DDTvar$. 
Integrated over a discrete time interval $\Delta t$,
$X(t + \Delta t) \sim \distNorm\left( X(t), \; \DDTvar \, \I \cdot \Delta
  t \right)$~\cite{ddt}.
Thus, latent state along the tree evolves according to collective
Brownian motion, where each branch event signifies the birth of two
independent Brownian motion processes (conditioned on their starting
location).

A draw from the distribution over $\nleaves$-leaf DDTs, marginalized over the
paths between nodes, consists of a set of locations (latent states and pseudotimes) for internal and leaf nodes, indexed by $\ibranch$:
\[
  \tau = \{(\lambda_\ibranch, t_\ibranch)\}_{\ibranch=1}^{2\nleaves-1}. 
\]
The canonical DDT model uses these locations to model a latent hierarchy
over data points, each generated from the latent state at a leaf (either
marginalizing over internal nodes or instantiating them for convenience
on the way to modeling the leaves)~\cite{ddt,ddt2}.

\subsection{Augmented Dirichlet diffusion trees for inferring cell trajectories} \label{sec:ourddt}

\figddt

We model observed expression profiles as arising from an underlying
branching process. In particular, we model the hidden abstraction of
cellular differentiation state, $\lambda \in \reals^{\ngenes}$ over
$\ngenes$ genes, as draws from a latent tree, and relate these smoothly
evolving latent vectors to the observations through an appropriate noise
model (detailed in \Cref{sec:obs_model}).
Next, we describe in more detail the generation of
$\lambda_\icell \defined [\lambda_\icell^{(1)}, \ldots, \lambda_\icell^{(\ngenes)}]$
for genes $1, \ldots, \ngenes$ per cell, $\icell$.

As classically formulated, the DDT specifies 
flexible densities for data generated at the leaves of a binary
tree~\cite{ddt} (\Cref{fig-ddt}, left). We extend the Dirichlet diffusion
tree to provide flexible densities that generate latent values (cell
states) according to a continuous-time distribution over the full tree
(\Cref{fig-ddt}, right).

Recall that a draw from the DDT process yields a set of locations at
discrete nodes, as well as a means of sampling all 
countably infinite locations between nodes (\Cref{sec:ddt}).
In practice, we do not need the full 
continuum of branch locations (gray lines, Figure~\ref{fig-ddt})
but rather focus on that of each cell (blue dots, right,
Figure~\ref{fig-ddt}).
Therefore, our algorithm instantiates the set of tree nodes, $\tree$, and
additionally instantiates only those branch locations that correspond to
cells. We can think of each location, or point, in the tree as a pair
comprising the state $\lambda$ (horizontal axis in \Cref{fig-ddt}) and
pseudotime $t$ (vertical axis). Thus, cell state $\lambda_\icell$ can be
seen as a projection of that cell's overall location in the tree,
$(\lambda_\icell, t_\icell)$.

In our case, the number of leaves (cell fates) is no longer fixed to the
number of data points---as in a classical DDT---so we
add a prior to regularize tree complexity,
e.g.\ $\nleaves \sim 1 + \distPoiss(\nleaves_0)$,
where $\nleaves$ corresponds to the number of particles that generate
the DDT (i.e.\ the number of leaves).
Additionally, we replace the canonical origin (typically set to the
$\ngenes$-dimensional zero vector at time $0$~\cite{ddt,ddt2}) with some
initial root location $(\mu_0, 0)$,
corresponding to a typical value for the data. Here, we leverage prior
knowledge about average expression profiles for stem cells (which can be
experimentally enriched and sequenced for this purpose, or readily identified
among a broader dataset).
To allow for gene-specific diffusion, we designate the diffusion
variance $\DDTvar$ to be a length $\ngenes$ vector and place a conjugate
inverse gamma prior over each component $\sigma_0^{2 \ (\igene)}$.

Thus, we model differentiation as the process by which latent state vector
$\mu_0$, characterizing
the starting population of stem cells, morphs and bifurcates to give rise to a
multiplicity of latent states.
These latent states, in turn, characterize
the diverse fates and functions of mature cells---noisily reflected by
their warped distributions over gene expression space.
We now have all of the ingredients to sample a set of tree nodes $\tree$
(yet to be populated by cells) according to an augmented DDT with leaf
prior $\nleaves_0$, concentration $\conc$, and origin $\mu_0$
(as well as $\DDTvar$ prior parameters, elided for
brevity)---which we denote $\distDDT(\nleaves_0, \conc, \mu_0)$.

Given $\tree \sim \distDDT(\nleaves_0, \conc, \mu_0)$, each cell $\icell$
diffuses down this tree, navigating branches according to a
\textit{rich-get-richer} scheme, until a random time point (detailed in
\Cref{app-ddt-ours}).
Specifically, we draw $t_\icell \sim \timedist(\cdot)$, where $\timedist$
is some distribution over $[0, 1]$ encoding our belief about how cells
are distributed over the tree. For example, we expect hematopoietic
samples to be skewed toward mature cells~\cite{hematopoiesis-spectrum},
so might choose $\timedist = \distBeta(4, 1)$.
The cell's location in the tree at this time yields a latent cell state
$\lambda_\icell \in \reals^{\ngenes}$ according to the Brownian bridge
defined by its Markov blanket. Namely, 
  \begin{equation} \lambda_\icell \sim \distNorm \left( \lambda_a +
      \frac{t_\icell - t_a}{t_b - t_a} (\lambda_b - \lambda_a) \; , \;\;
      (t_\icell-t_a) \left( 1 - \frac{t_\icell - t_a}{t_b - t_a} \right) \DDTvar
      \, \I \right),
  \end{equation}
for (cell or node) locations
$\{ (\lambda_a, t_a), (\lambda_b, t_b) \mid t_a < t_b \}$
bookending cell $\icell$ (see \Cref{app-ddt} for more details about how
the Brownian bridge formulation follows naturally from the properties of
the DDT).
Finally, we sample the cell's observed 
expression profile
$x_\icell \in \nats^{\ngenes}$ given $\lambda_\icell$
according to some noise 
model, defined below (\Cref{sec:obs_model}).
Here, we make the simplifying assumption that genes are expressed
independently conditioned on the latent 
structure of differentiation.
Importantly, this generation scheme captures the desired behavior that
we can model vastly different probabilities of observing a cell along
certain intervals (not only specific time intervals, but also specific
branches)---accurately reflecting the biological variance in
differentiation ``velocities'' and in the popularity of various cell
fates.

\subsection{Observation model for single-cell RNA-seq} \label{sec:obs_model}

Now we need a way to connect the noisy scRNA-seq cell snapshots we observe to
our tree-structured abstraction over the latent dynamics of cell state.
Consider a single cell containing a set of messenger RNA (mRNA) transcripts
corresponding to each gene that is currently expressed---some subset of all
$\ngenes$ genes in its genome. We posit that, for a 
particular cell type, the discrete count $\ntranscripts$ of transcripts of a
particular gene $\igene$ has a Poisson distribution with rate
$\lambda^{(\igene)}$:
$
\ntranscripts \sim \distPoiss(\lambda^{(\igene)}).
$

Throughout the workflow for droplet-based sequencing~\cite{DropSeq, droplet,
10X}, the current state of the art for single-cell transcriptomics, there are
several processes known to affect accurate observation of
$\ntranscripts$~\cite{sc-rev-regev, sc-rev-teichmann, sc-rev-tech-bio}
(\Cref{app-obs}). In brief, transcripts must hybridize (with probability
$\pbead$) to a primer containing an \iid\ uniform primer-specific barcode
(unique molecular identifier, or UMI)~\cite{DropSeq}, must be amplified
(w.p.\
$\pamplify$) over the course of $\rounds$ rounds of PCR, and must hybridize to
the flow cell for sequencing (w.p.\ $\pseq$).
Since the original quantity $\ntranscripts$ was Poisson distributed, we can use
the thinning property and the marking property of Poisson processes to show that
the number attached to each unique UMI and ultimately sequenced is
\[
  \ntranscripts_1, \dots, \ntranscripts_{\numi} \distiid
  \distPoiss\left(\frac{(1+\pamplify)^\rounds \, \pseq \, \pbead \, \lambda^{(\igene)}}{\numi}\right),
\]
where $\numi$ is the theoretical number of distinct UMIs (i.e., $4^{\text{\#basepairs}}$).
Finally, transcripts are quantified by aligning sequences to a reference
genome, resulting in an overall count for this particular gene of
$
  x^{(\igene)} = \sum_{i=1}^{\numi} \ind\left[ \ntranscripts_i > 0 \right].
$
The distribution of this quantity has a closed-form expression:
\[
  x^{(\igene)} \dist \distBinom\left(
    \numi \; , \;\;
    1 - e^{-\dropout \lambda^{(\igene)}}
  \right) \qquad \textrm{ with } \label{eq:observationModel}
  \qquad
  \dropout \defined \frac{(1+\pamplify)^\rounds \, \pseq \, \pbead}{\numi},
\]
where $\dropout$ is a hyperparameter accounting for gene dropout (optionally,
$\dropout^{(\igene)}$, with gene index $\igene$ also superscripting the
probabilities on the right). Due to experimental dropout and natural gene
regulation, the observed expression profile for cell $\icell$,
$x_\icell \defined [x_\icell^{(1)}, \ldots, x_\icell^{(\ngenes)}]$,
is a sparse vector of digital counts with $G \approx 20,000$
(for human cells)~\cite{sc-rev-regev, sc-rev-teichmann, sc-rev-tech-bio}.
In practice we model a subset of the most variable genes, since genes
with low variance over sampled profiles
contain little information to resolve cell states along a trajectory.

Notably, latent state $\lambda_\icell$ drawn from a tree is
in $\reals^\ngenes$ (\Cref{sec:ourddt}), whereas the vector of Poisson
rate parameters for gene expression must be nonnegative.
So, we replace the original Poisson parameter per gene 
with $\link(\lambda^{(\igene)})$ for some link function $\link: \reals
\to \reals_{\geq 0}$. This maneuver enables us to model states along the
tree as unconstrained reals, while preserving the necessary sign in the
observation model (\Cref{eq:observationModel}).
For efficient inference, we derive a link function that
permits P\'olya-gamma augmentation~\cite{pg} for conditional conjugacy
(elaborated in \Cref{sec:inference} and \Cref{app-pg}).

Conveniently, since cell state $\lambda_\icell$ drawn from a diffusion tree is conditionally Gaussian, 
the dropout parameter $\dropout^{(\igene)}$ (which acts as a
per-gene linear scalar in the observation model) 
can be absorbed by the diffusion variance $\DDTvar$ and therefore learned. 

\newcommand{\infitem}[1]{\textbf{#1}:\ }

\section{Inference} \label{sec:inference}

Having specified the generative model, we derive a Markov chain Monte
Carlo sampler to approximate the Bayesian posterior over parameters of
interest (as in \Cref{sec:bayes}).
Specifically, we aim to recover tree topology and cell locations (rates,
pseudotimes, and branches)---collectively, $\theta$---given the observed cell profiles---$x$.
Here, we describe our inference algorithm for efficient mixing over
trees, beginning with the design of Metropolis-Hastings proposals for
structure-learning and ending with a message-passing framework for exact
belief propagation of latent cell and node states.
In practice, we cycle through these steps uniformly at random but linger
on cell resampling for enough iterations such that every cell is
resampled in expectation (per cluster of cell resampling moves).
See~\Cref{app-figs}, \Cref{fig-inference} for a graphical summary of the
algorithm.

\infitem{Cell resampling}
    In the original DDT model, each data point is always fixed to a
    particular leaf, and latent states are resampled during inference by
    rearranging the underlying tree structure.
    However, we face the daunting challenge of determining where each cell belongs
    on the tree---resampling time and branch assignment, in addition
    to latent state (while simultaneously learning the tree itself).
    To this end, this step proposes new locations for a
    random subset of cells, conditioned on $\tree$. This is akin
    to a Gaussian mixture model over extant branches at each time slice
    $t_\icell$ and $t_\icell'$.
    We design proposals where cells select branches with probability
    proportional to their likelihood,
    so the Metropolis-Hastings acceptance ratio reduces to the
    (log sum over)
    categorical normalizing constants at each time slice.

    Motivated by empirical analysis of trace plots,
    we accelerate mixing by making proposals that simultaneously resample
    all cells within randomly sized time partitions, 
    interleaved with occasional proposals that simultaneously 
    resample cells on a single branch.
    Intuitively, this strategy is advantageous because Brownian motion of
    cell states at a given time is dependent on the placement of
    neighboring cells on each branch. Thus,
    we are doomed to 
    slim acceptance of proposals unless we give an entire region of cells
    the opportunity to switch branches (in random order). We also
    increase acceptance of cell proposals by using a variable augmentation
    scheme to sample new cell states (see message passing, below).

    While a proposal of this sort is not strictly necessary for inference
    on the original DDT model,
    a modification of this move could improve mixing
    by jumping between states that would otherwise be separated by many
    moves, potentially with unlikely intermediate transitions.
    Specifically, tailored for the original DDT model,
    this step would entail proposing ``leaf swaps''
    (akin to cell resampling at fixed time $1$, but requiring a 1:1
    mapping between leaves and data points at the end of each move).

\infitem{Subtree prune and regraft (SPR)}
    This proposal, as formulated for DDTs~\cite{ddt,ddt2},
    detaches a random subtree and draws its new parent from the prior.
    However, unlike in Neal's original formulation, we need a way to deal
    with the cells that fell along the deleted branch, while accounting
    for effects on the likelihood of neighboring cells.
    We supplement this
    move by resampling cells over the affected subtree, which is defined
    by the most recent common ancestor (MRCA) of the original and
    proposed parent nodes. 
    By confining cell resampling to the MRCA subtree,
    we isolate changes to the likelihood to this local subtree
    (which would otherwise be induced through global effects on
    self-reinforcing cell paths).

\infitem{Split/merge}
    Whereas the original DDT pinned each data point to a leaf (therefore
    fixing the complexity), we need a way to grow and shrink tree depth
    according to the data.
    This step proposes growing or pruning subtrees, changing the dimension of the
    tree by one or more leaves (and resampling affected cells).
    ``Split''
    chooses a random subtree to collapse, whereas ``merge'' chooses a
    random number of leaves to generate $\nleaves'$, and draws from the DDT
    prior to generate a new $\nleaves'$-leaf subtree that diverges from a
    random existing branch. As in SPR, we resample cells only over the
    relevant subtree to preserve cell likelihoods on the remainder of the
    tree.

\infitem{Gibbs sampling}
    We perform Gibbs updates to
    diffusion variance $\DDTvar$ (with a conjugate inverse
    gamma prior for Gaussians with fixed mean and unknown variance), as
    originally suggested by Neal~\cite{ddt,ddt2}.
    Here, the posterior over $\DDTvar$ (a global scalar, or vectorized
    over genes) is updated based on the relative locations of all cells
    and tree nodes.

\infitem{Message passing}
    We derive message passing to perform exact inference on node and cell
    states given their times and topology.
    Here, we leverage
    belief propagation on a binary tree with Gaussian emissions, made
    conditionally conjugate through P\'olya-gamma (PG)
    augmentation~\cite{pg} of binomial observations.

    The PG trick has previously been used for conjugacy in binomial
    models like logistic regression~\cite{pg},
    and has been adopted to 
    facilitate computation
    for non-Gaussian
    potentials within a message-passing framework~\cite{slds}.
    However, this work represents a novel synthesis with the more general
    algorithm for Gaussian belief propagation, originally posited for
    solving systems of linear equations through a probabilistic
    lens~\cite{gbp}.

    This augmentation scheme is made possible by the choice of an
    appropriate link function $\link(\cdot)$ 
    such that we can rewrite the original 
    observation model, 
    \[
      x \mid \lambda \sim \distBinom\left( \numi,\;
        1 - \exp\left[-\dropout \cdot \link(\lambda) \right] \right),\label{eq:binom1}
    \]
    in the form amenable for the P\'olya-gamma trick,
    \[
      x \mid \pgaug \sim \distBinom\left(N ,\;
        \frac{1}{1 + \exp(-\pgaug)} \right),\label{eq:binom2}
    \]
    for some variable $\pgaug$ with prior $\distNorm(\mu, \Sigma)$.
    This setup is desirable because it allows us to 
    elide the binomial likelihood in favor of more tractable distributions, by
    alternatingly sampling
    \[
      \omega \mid \pgaug \sim \distPG(N, \pgaug)
    \qquad \text{and} \qquad
      \pgaug \mid x, \omega \sim \distNorm(m_\omega, V_\omega), \label{eq:algorithm}
    \]
    where
    \[
    V_\omega = (\Omega + \Sigma\inv)\inv,
    \qquad
    m_\omega = V_\omega(\Sigma\inv \mu + x - N/2),
    \]
    and $\Omega \defined \diag(\omega)$~\cite{pg}. 
    Importantly, the PG density admits 
    efficient sampling~\cite{pg}.

    To exploit this scheme, we choose 
    link function $\link$ such that the binomial probability term in
    \eq{binom1} is equivalent to the sigmoidal probability in \eq{binom2},
    for $\pgaug = \dropout \, \lambda$. Namely, $\link(\cdot)$ must be
    \[
      -\log\left( 1 - \frac{1}{1 +
          \exp(-\dropout \, \lambda)} \right) / \dropout.
    \]
    Note that the link function now depends on the dropout parameter. Since this reparameterization
    effectively scales a Gaussian random variable ($\lambda$) by a scalar
    ($\dropout$), the dropout parameter can be absorbed by the diffusion variance
    $\DDTvar$ (and therefore learned, per gene or per tree).
    That is, we directly model the scaled proto-rates $\dropout\,\lambda$  as
    latent variables sampled along the tree, rather than explicitly setting
    $\dropout$ and modeling its effects.

    Now, we endow each cell with an auxiliary PG variate $\omega_\icell \in
    \reals^{\ngenes}$. Completing the augmentation scheme, we let
    $N=\numi$ and let $\mu, \Sigma$ be the prior over $\lambda_\icell$
    given Gaussian diffusion on the tree;
    then,
    \eq{algorithm} is a valid algorithm for efficiently resampling latent
    cell states while leaving cell likelihoods intact.

    Both cells and DDT nodes are now conditionally Gaussian and form a directed
    acyclic graph, so we can perform continuous belief propagation~\cite{gbp} to
    ``share'' information across the full tree and exactly resample all points
    (cells and nodes), recursively conditioned on their respective neighbors. This
    algorithm requires only two (subtree-parallelizable) sweeps over the tree---once
    to pass messages upward from leaves to root, and once to pass recomputed
    messages downward from root to leaves---prior to 
    simultaneous resampling of all latent states (\Cref{app-pg}).

    This move supplants Neal's original suggestion of resampling latent
    states associated with internal nodes via Gibbs updates (conditional
    on the locations of their immediate neighbors)~\cite{ddt,ddt2}.
    Gaussian belief propagation (with PG augmentation, if data generation
    from latent states at the leaves is nonconjugate) presents a valid
    alternative for DDTs.

    Notably, Knowles \emph{et al.}~\cite{varddt} derive
    a variational message passing framework for DDTs, to
    speed up calculation of (approximate) evidence
    and drive efficient search over tree structures,
    but their algorithm and intention are disjoint from ours.
    Later, the same authors use belief propagation to marginalize over
    internal nodes in the DDT~\cite{pydt}, 
    but this effort is also distinct and
    they do not consider the case of message passing with non-Gaussian
    observations.

\subsection{Initialization} \label{sec:init}

While MCMC is guaranteed to reach equilibrium at its stationary distribution,
manufactured to be the true posterior,
we can accelerate equilibration of our finite-time approximate inference
algorithm by careful initialization. We initialize the sampler by drawing
a tree from the DDT prior.
Then, we initialize cell locations by sampling latent cell states
conditioned on their corresponding data and auxiliary PG variates for
each branch (at some time $t$, sampled from the time distribution or
given by experiment).
Finally, from these proposals, we iteratively assign
cells to branches with
probability proportional to their likelihoods. This strategy is
equivalent to a cell resampling proposal over the full tree.

\newcommand{\rfigwidth}{0.574\textwidth}

\newcommand{\rfigs}{
\begin{figure}
  \vspace{24pt}
    \centering
    \raisebox{0pt}[\dimexpr\height-4.1\baselineskip\relax]{
      \includegraphics[ width=\rfigwidth ]{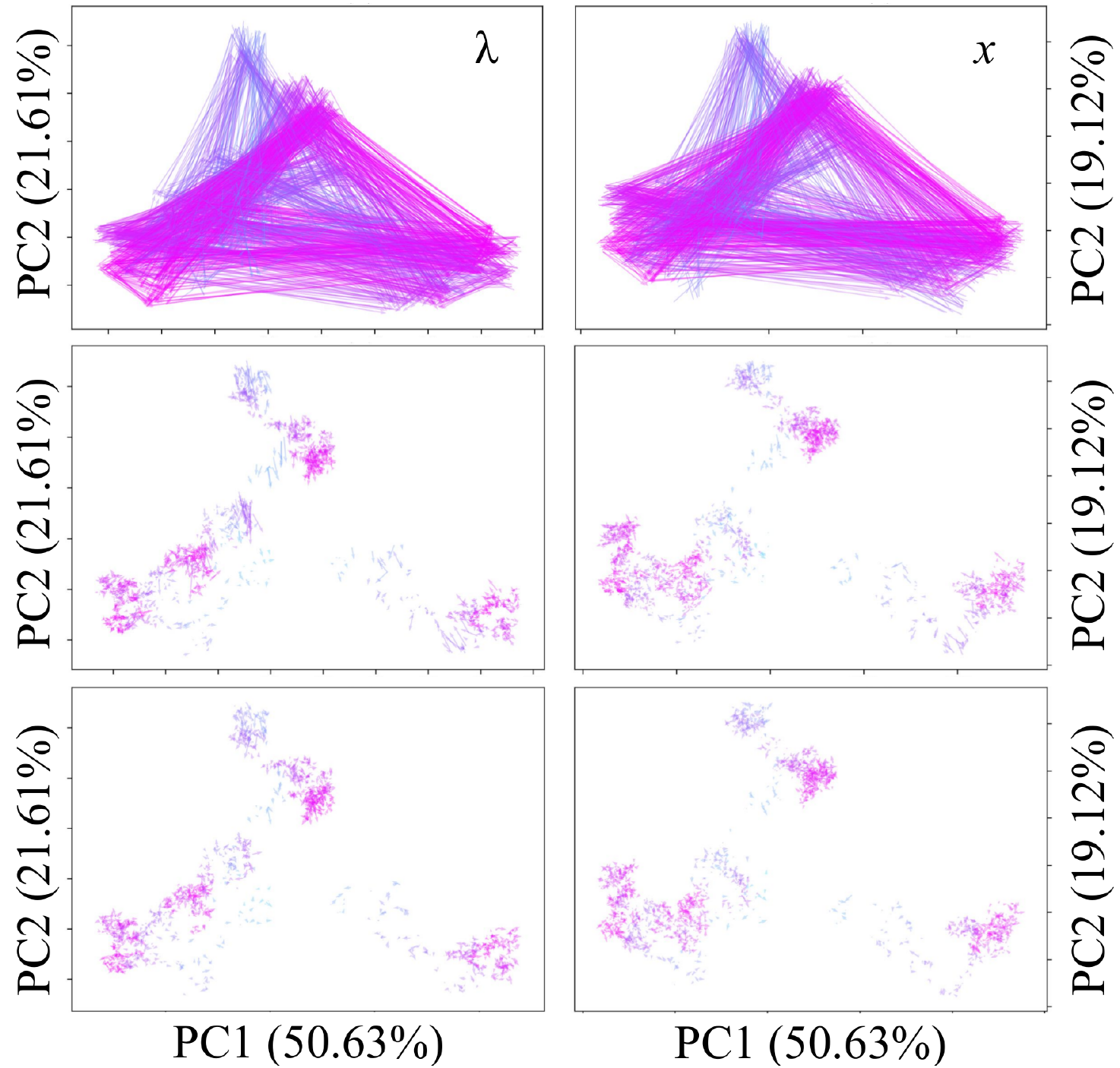}
    }
    \caption{Shifts in $\lambda$ (\emph{left}) and $x$ (\emph{right})
        along the first two principal components (PCs).
        Arrows, each corresponding to a single cell, are colored by true
        time and point from true to inferred value.
        Percentages on axis labels indicate the
        portion of total variation
        explained by that axis (i.e., that PC).
      \emph{Top to bottom}:
      random, initial, MAP trees.}
    \label{fig-shifts}
    \vspace{0.6em}
    \includegraphics[ width=\rfigwidth ]{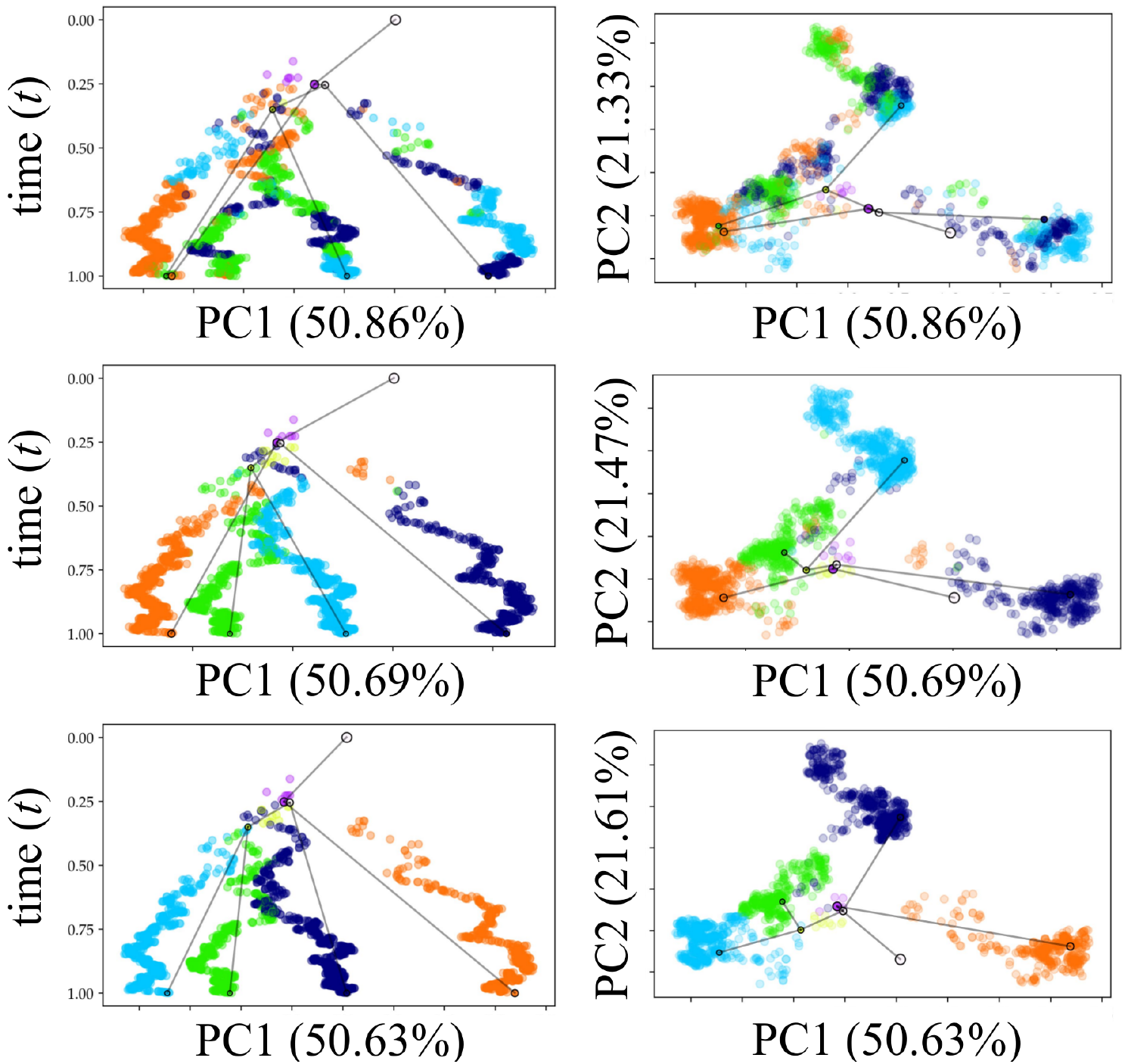}
    \caption{Cells colored by branch assignment.
    \emph{Left:} PC1 of cell rate vs.\ time. \emph{Right:} PC1 vs.\ PC2.
    \emph{Top to bottom}: initial, MAP, true trees.}
    \label{fig-branches}
    \vspace{0.6em}
    \includegraphics[ width=\rfigwidth ]{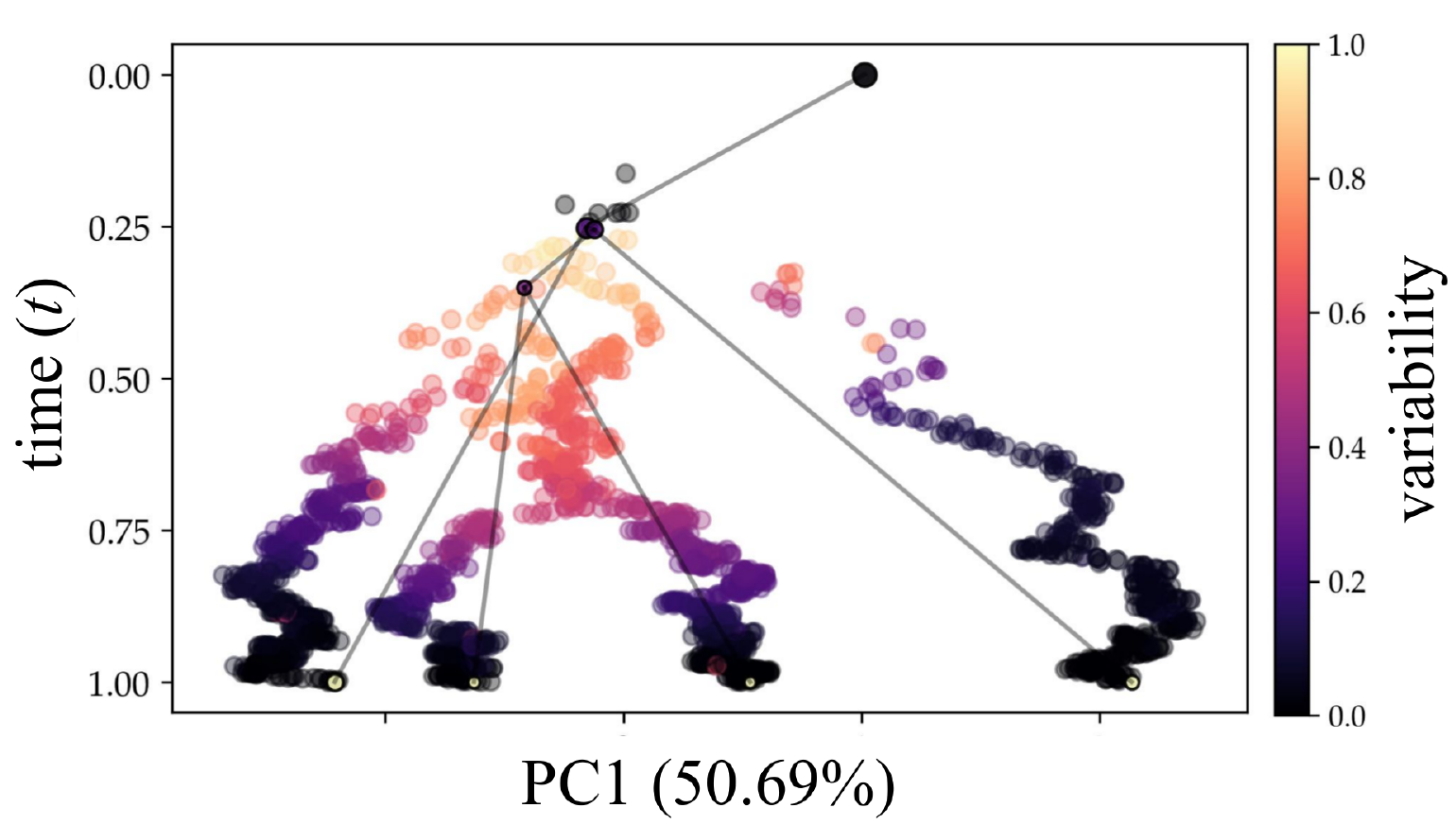}
    \caption{Relative branch variability per cell,
    based on categorical dispersion~\cite{dispersion} 
    across samples
    (normalized by choices per time slice
    and scaled 0-1), plotted atop the MAP tree.}
    \label{fig-variability}
\end{figure}
}

\section{Experiments} \label{sec:results}

\rfigs

In preliminary experiments, we simulated single-cell data to verify
inferred parameters against known ground truth.
Focusing on a restricted model 
(fixing tree topology and
cell/node times), we demonstrate visually and quantitatively that our
algorithm recovers latent trajectories.
Notably, this fixed-time regime is applicable to time-course
experiments where cell times are (approximately) known and
tree topology is given by prior knowledge.

We simulated data for 2000 cells by sampling times
$t_\icell \sim \distBeta(4, 1)$ and
drawing latent rates $\lambda_\icell \in \reals^{10}$ from an augmented DDT
with 4 leaves and concentration $\alpha=3$. We then simulated
expression profiles $x_\icell \mid \lambda_\icell$, as in
\Cref{sec:obs_model} and \Cref{app-pg}. 

Following 9050 iterations of MCMC, we examined sampled trees
(thinning=50) and assessed convergence based on trace plots.
We largely recover cell parameters 
through our initialization procedure alone (\Cref{fig-shifts}; procedure detailed in \Cref{sec:init}).
Specifically, we highlight the discrepancy between true and inferred
parameters per cell as the length of the arrow between them
(visualized following PCA).
We observe that cells' deviation from ground-truth is miniscule for
sampled trees in comparison to a random tree drawn with identical
parameters (\Cref{fig-shifts}, top).
These arrows shrink further from initialization
to the maximum \emph{a posteriori} (MAP) tree\footnote{More
precisely, the sampled tree with highest probability.}
(\Cref{fig-shifts}, middle and lower, respectively),
indicating that inference is approaching ground truth.
We also show recovery of cell branch assignments (\Cref{fig-branches}),
though we observe label switching.
This challenge of identifiability is ubiquitous to MCMC inference of
models involving clustering or other labeled components, and a
sizable literature has sprung up to address it (e.g.,\ 
purposeful initialization near a posterior mode)~\cite{labels,stan}.

To quantitatively assess tree recovery, or even summarize the posterior beyond a single
mode, we need a similarity score for differentiation trees. 
Motivated by the desire for a means of comparing trees that may be of
differing depths, such that node-matching is challenging or impossible,
we propose a ``triplet metric'' (\Cref{app-triplet}). This similarity
score abstracts away the underlying tree to prioritize the topology of
the cells themselves and is agnostic to tree size.
Further, the triplet metric is invariant to label switching.
Using the triplet
metric, we underscore our visual recovery of cell branches
(\Cref{fig-branches}) by
quantifying how well cell relationships agree with ground truth.
Specifically, we randomly subsample triplets of cells and compare their
relative locations along the inferred and true trees.
Within a given triplet, we compute pairwise cell distances 
to determine the outlier cell on each tree (or its complement, the closest pair),
where distance is given by branch length (in $t$) required to traverse
the tree from one cell to the other (via their MRCA node, if not on the
same branch). Then, we compute the triplet metric as the proportion of
cell triplets whose outlier cell is consistent across both trees
(e.g., ground-truth and sampled, or between two samples).
To our knowledge, this simple method, which uses a distance metric
inspired by phylogenetic trees~\cite{phylo}, represents a new
contribution to the nascent literature on comparing 
cell fate trajectories~\cite{sc-compare}.
Here, $0$ 
corresponds to 
orthogonal cell topologies,
and $1$ corresponds to perfect concurrence with ground truth.
This score 
increased from
$0.520$ to $0.828$ from the initial to MAP tree (versus 
$0.409$ for the random tree), indicating convergence toward true
cell topology.

Finally, we can begin to reap the benefits of our probabilistic approach
by examining branch uncertainty per cell across sampled trees
(\Cref{fig-variability}), as quantified by a metric for categorical
variability~\cite{dispersion}. As expected, cells are most ``confident''
in their branch assignment near leaves (especially leaves that diverge
earlier and are more separated in latent space), and are least certain
where internal branches of the tree are close together or overlap.

\section{Conclusions and future} \label{sec:conclusion}

We take a Bayesian nonparametric approach to learning cell state from inherently
noisy single-cell RNA-seq data of differentiating cells. 
In particular, we define cell state by the latent parameterization of a
distribution over gene expression space, and model these latent vectors as
arising from bifurcating, self-reinforcing paths along a probabilistic tree.
Motivated by the biology, this approach necessitated the development of a novel
probabilistic model for data generated continuously over a tree, rather than
restricted to discrete nodes or leaves.
Performing efficient inference on this model also required innovations:
we develop a message-passing framework for exact Gaussian belief propagation on
a binary tree through P\'olya Gamma augmentation and a novel MCMC
sampler for learning tree structure.
Simulation experiments show that our
inference algorithm
can resolve latent trajectories from single-cell data.
Further, we demonstrate the benefits of an interpretable latent
structure that includes 
a coherent framework for uncertainties over
parameters like cell branch locations, rather than assigning each cell a
single definitive label.

We are working to apply this model to data sets with noisy time-labels
(e.g.,\ experimental time points of zebrafish embryos
post-fertilization), which we can use to helpfully constrain
inference---as well as data sets from the more challenging regime of
naturally mixed, asynchronous cells
(e.g.,\ hematopoietic, tracheal, and gut epithelial cells).
We can then validate and refine the model through
(experimental or natural)
perturbation.
By targeted experimental perturbation~\cite{perturb-seq},
e.g.\ of genes inferred as essential to a particular branch decision,
we can interrogate predictions about the 
map from expression to tree topology.
On the other hand, natural perturbations such as disease present an opportunity
to compare healthy and aberrant 
topologies, 
and infer the driving discrepancies
at the molecular (gene regulatory) level.

We are also developing an extension of the model that allows for a ``trunk'' of
potential trajectories between branches (modeled as switching linear dynamical systems),
such that branch points are hyperplanes (rather than zero-dimensional) and cells
can approach from many distinct paths. This more complex model aims to
decode cell fate decisions, by
discriminating between trajectories (and their driving expression
dynamics) of distinct sub-populations of cells prior to bifurcation
events.

Ultimately, we are interested in the stochasticity of lineage fate
specification---whether cells ``commit'' to (are probabilistically inclined
toward) particular fates prior to branching---and in reconstructing the master
regulatory programs and fitness landscapes that govern large-scale changes in
cell state.

\clearpage
\small
\printbibliography

\clearpage
\appendix
\beginsupplement

\section{Supplemental figures}\label{app-figs}

\begin{figure}[H]
  \centering
  \includegraphics[width=.9\textwidth]{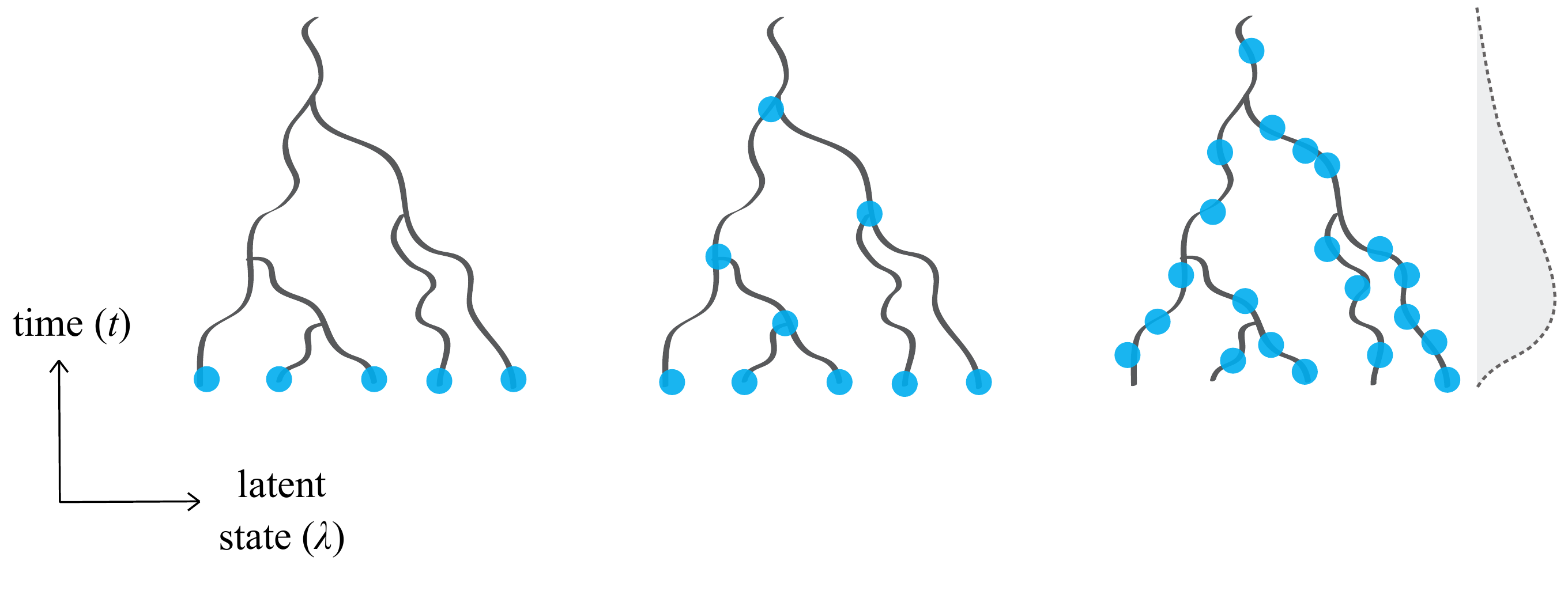}
    \caption{
    Categories of probabilistic latent tree-structured models. Existing
    Bayesian tree models (\emph{left \& middle}) did not fit our desiderata
    for observations generated continuously over a latent tree, rather than
    confined to discrete nodes, so we derive a novel tree-structured
    generative process (\emph{right}). Concretely, from left to right, we
    depict (1) ``latent ancestry'' models, where data is generated only at
    the leaves
    (e.g.\ Dirichlet~\cite{ddt,ddt2} or Pitman--Yor~\cite{pydt} diffusion
    trees, hierarchical Dirichlet process~\cite{hdp}, Kingman's
    coalescent~\cite{coalescent}), (2) hierarchical clustering models, where
    data is generated at the nodes of the tree (nested Chinese restaurant
    process~\cite{nCRP}, tree-structured stick breaking process~\cite{tssb}), and (3) our augmented Dirichlet diffusion tree model,
    which---in contrast---admits a continuum of latent states over the full tree according to
    a continuous time distribution on the unit interval (\emph{inset far right}).
      }
    \label{fig-3trees}
\end{figure}

\vspace{0.6em}

\begin{figure}[H] 
  \centering
  \includegraphics[width=.6\textwidth]{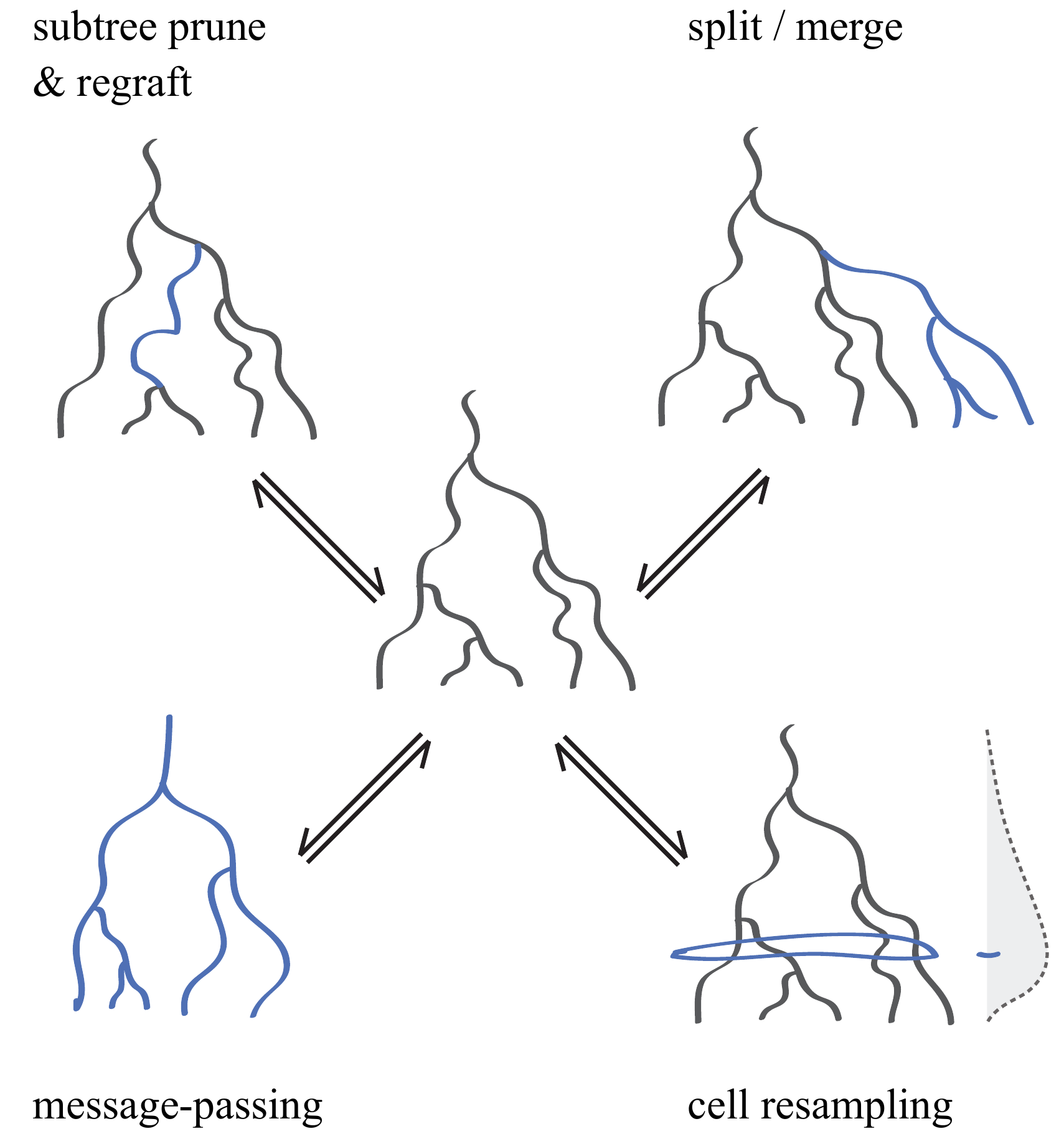}
    \caption{
      Cartoon overview of moves made by our inference algorithm.
      \emph{Upper}: Moves that resample latent topology---changing internal
      structure (subtree prune \& regraft) or the size of the tree
      (split/merge).
      \emph{Lower}: Moves that resample latent locations---either cell and node
      latent states (message passing), or cell times and latent states (cell
      resampling).
      }
    \label{fig-inference}
\end{figure}

\section{Dirichlet diffusion trees}\label{app-ddt}

The Dirichlet diffusion tree (DDT) model provides a family of priors over
infinitely exchangeable data that derive from a latent binary branching process.
As classically formulated, the DDT generalizes Dirichlet process mixture models
for data that are hierarchical, such that data points are generated at the
leaves, and internal nodes correspond to hierarchical clusters~\cite{ddt}.
Alternately, the DDT can be constructed as the continuum limit of the
nested Chinese restaurant process, or as the dual of the Kingman's
coalescent~\cite{knowles-thesis}.

Consider a draw from the distribution over $\nleaves$-leaf DDTs. If marginalized over
the paths between nodes, the sampled DDT consists of a set of discrete
node locations (latent states and pseudotimes)
\[
  \tau = \{(\lambda_\ibranch, t_\ibranch)\}_{\ibranch=1}^{2\nleaves-1}.
\]
Here, we use $\ibranch$ to index the locations of tree nodes (internal nodes
and leaves), whereas we later use $\icell$ to index the locations of cells
along the tree.

Tree topology is generated by iteratively simulating paths of particles
according to a Wiener process of Gaussian diffusion (i.e.\ Brownian
motion). Specifically,
a particle that has reached $X(t)$ at time $t \in (0, 1)$ will diffuse to $X(t +
dt) = X(t) + \distNorm(0, \; \DDTvar \, \I \cdot dt)$ after an infinitesimal
amount of time $dt$, for some $\DDTvar$ that governs diffusion (which can be
learned). Integrated over a discrete time interval $\Delta t$, then,
$X(t + \Delta t) \sim \distNorm\left( X(t), \; \DDTvar \, \I (\Delta t) \right)$~\cite{ddt}.

Following Neal~\cite{ddt}, we assume a probabilistic branching rate of
$\branchrate(t) = \conc / (1-t)$, where $\conc$ is a smoothness parameter
related to whether branches are concentrated toward the root or the leaves. Let
\[
  \cbranchrate(t) \defined \int_0^t \branchrate(u) \d u = -\conc\log(1-t);
\]
this is the cumulative branching function. If a particle is on a leg of the tree
bookended by times $[t_a, t_b]$, and there are $\nparticles$ particles that have
traversed this path, the probability of branching at some time $t \in (t_a,
t_b)$ is
\[
  \pbranchonleg_{t_a}(t) \defined \Pr(\mathrm{branch \ in } \ [t_a, t]) = 1 -
  e^{(\cbranchrate(t_a)-\cbranchrate(t)) / \nparticles} \equiv 1 -
  \left(\frac{1-t}{1-t_a}\right)^{\conc/\nparticles}.
\]
To see when/if a new particle branches on an existing leg of the tree $[t_a,
t_b]$, we calculate $t_\ibranch$ by the inverse CDF method. If $t_\ibranch >
t_b$, then the particle does not create a new branch on this leg and instead
follows one of the existing branches at time $t_b$.

Overall, to create a DDT with $\nleaves$ particles:
\begin{enumerate}
\item Set the root of the tree to some origin $(\mu_0, 0)$, corresponding to a
  typical value for the data---here, we leverage prior knowledge about average
  expression profiles for stem cells. Draw the first particle's leaf location as
  $\lambda_{\ell_1} \sim \distNorm(\mu_0, \; \DDTvar \, \I)$. \item For $\ileaf = 2, \ldots, \nleaves$:
  \begin{enumerate}[label=\alph*)]
  \item Find the next branch point and time, $(\lambda_b, t_b)$, (e.g.\ for
    $\ileaf=3$, this will be where the second particle diverged) and find the number
    of particles $\nparticles$ that have taken this path (equivalently, the
    number of leaves under node $b$). Draw $u \sim \distUnif(0,1)$ and compute a
    proposed branching time,
    $$ t_\ibranch = \pbranchonleg\inv_{t_a} (u) = 1 - e^{\log(1 - t_a) + \frac{\nparticles}{\conc} \log (1 - u)}. $$
  \item If $t_\ibranch < t_b$, branch at time $t_\ibranch$. Sample the branching
    location according to the Brownian bridge defined by its Markov blanket, i.e.\
    nodes $a$ and $b$ bookending the start and end of this leg,
    \begin{equation} \lambda_\ibranch \sim \distNorm \left( \lambda_a +
        \frac{t_\ibranch - t_a}{t_b - t_a} (\lambda_b - \lambda_a) \; , \;\;
        (t_\ibranch-t_a) \left( 1 - \frac{t_\ibranch - t_a}{t_b - t_a} \right) \DDTvar
        \, \I \right). \label{eq:brownianBridge}
    \end{equation}
    This equation derives from the properties of Brownian bridges, which
    describe Brownian motion 
    tamped to prescribed values at both ends. A Brownian
    bridge on the interval $[0,T]$, starting at $X_0 = 0$, and ending at $X_T = 0$
    has $X_t \sim \distNorm(0, \; t(1 - t/T) \DDTvar \, \I)$. We want a bridge
    between $\lambda_a$ and $\lambda_b$ for times $t_\ibranch \in (t_a, t_b)$, or,
    equivalently, $t_\ibranch - t_a \in (0, t_b - t_a)$. Rewriting the time interval
    this way yields the variance in \eq{brownianBridge}, and the mean comes from
    interpolating between $\lambda_a$ and $\lambda_b$ over time. Record this branch
    point $(\lambda_\ibranch, t_\ibranch)$ and sample its final leaf location from
    $\distNorm(\lambda_\ibranch, (1-t_\ibranch) \, \DDTvar \, \I)$. Move on to the
    next particle.
  \item If $t_\ibranch > t_b$, do not branch off of this leg. Instead, pick one
    of the two branches at time $t_b$ with probability equal to $\nparticles_i /
    \nparticles$, where $\nparticles_i, \; i \in \{1,2\}$ is the number of particles
    that previously chose that branch. Go back to step a).
    \end{enumerate}
\end{enumerate}

In order to model gene-specific diffusion, we let the diffusion variance
$\DDTvar$ be a length $\ngenes$ vector and place a conjugate inverse gamma
prior over each component $\sigma_0^{2 \ (\igene)}$. 

\section{Model for latent cell states conditional on a tree}\label{app-ddt-ours}

A draw from the DDT process provides a set of locations at internal nodes and
leaves, as well as a means of sampling all diffusive locations between nodes.
In order to model cells as arising from a continuous-time distribution
over a Dirichlet diffusion tree, we draw $\tau \mid \nleaves \sim
\distDDT$ and sample each cell $\icell$ as follows:
\begin{enumerate}
\item Draw $t_\icell \sim \timedist(\cdot)$, where $\timedist$ is some
    distribution over $[0, 1]$ that represents our belief of how cells are
    distributed over the tree---e.g.\ for hematopoiesis we expect most cells to be
    near the leaves~\cite{hematopoiesis-spectrum}, so might choose something like
    $\timedist = \distBeta(4, 1)$.
\item Traverse the tree, starting at the root ($\lambda=\mu_0, t=0$). At each
    branch point, select a branch with probability proportional to the number of
    cells that previously chose that branch (with counts offset by 1 for numerical
stability). Stop at time $t_\icell$.
\item Find the points on the chosen branch (nodes or cells) that have $t_a < t_\icell < t_b$,
    with no other points in between; let $\lambda_a$ and $\lambda_b$ be the latent states of these points.
    For example, if this is the first cell to be added to an ``empty'' branch,
    these values are the locations of the node and parent node that define the branch.
    Then, sample $\lambda_\icell$ according to the Brownian bridge defined by its
    Markov blanket (points $a$ and $b$), as in~\eq{brownianBridge}:
  \begin{equation} \lambda_\icell \sim \distNorm \left( \lambda_a +
      \frac{t_\icell - t_a}{t_b - t_a} (\lambda_b - \lambda_a) \; , \;\;
      (t_\icell-t_a) \left( 1 - \frac{t_\icell - t_a}{t_b - t_a} \right) \DDTvar \, \I
    \right).
  \end{equation}
\item Finally, for each gene $\igene \in \{1, \ldots, \ngenes\}$, sample gene
  expression level
  \[
    x_{\icell}^{(\igene)} \sim \distBinom\bigg( \numi \; , \;\; 1 - e^{-\dropout
      \, \link\left(\lambda_\icell^{(\igene)}\right)} \bigg)
  \]
  as in \eq{observationModel}, for positive link function $\link$.
\end{enumerate}

Completing the generative model, we also place a regularizing prior over
tree complexity, i.e.,\ number of leaves $\nleaves \sim 1 +
\distPoiss(\nleaves_0)$.

\section{Observation model for single-cell RNA-seq}\label{app-obs}

Consider a single cell containing a set of messenger RNA (mRNA)
transcripts corresponding to each gene that is currently expressed. We
assume that, for a particular cell type, the discrete count
$\ntranscripts$ of transcripts of a particular gene $\igene$ has a
Poisson distribution with rate $\lambda^{(\igene)}$,
\[
  \ntranscripts \sim \distPoiss(\lambda^{(\igene)}).
\]

For droplet-based methods, the current state-of-the-art for scRNA-seq,
cells are flowed through a microfluidic device such that each cell is
captured by a droplet of fluid containing a single microbead covered in a
large number (roughly $10^8$) of barcoded DNA primers~\cite{DropSeq,
droplet, 10X, sc-rev-tech-bio}. Each primer contains a PCR handle,
bead-specific barcode, and primer-specific barcode (unique molecular
identifier, or UMI),
as well as a poly-T tail designed to capture the 3' poly-A tail of the
processed mRNA molecules.
Following cell lysis, mRNA transcripts hybridize to these randomized
primers. Under the assumption that each droplet contains a single
microbead and a single cell, the bead-specific barcode acts as a
cell-specific barcode and the UMI (in tandem with the attached gene sequence)
acts as a transcript-specific barcode~\cite{DropSeq}. As there are vastly more
primers on the microbead than mRNA levels in the cell (at most roughly $10^6$),
we assume that each transcript hybridizes with probability $\pbead$ to a primer
with an \iid uniform UMI. Since the original quantity $\ntranscripts$ was
Poisson distributed, we can use the thinning property and the marking property
to show that the number attached to each unique UMI is
 \[
   \ntranscripts_1, \dots, \ntranscripts_\numi \distiid \distPoiss \left(\frac{\pbead\lambda^{(\igene)}}{\numi}\right).
\]
Ideally, each $\ntranscripts_i \in \{0, 1\}$; if $\ntranscripts_i > 1$, we will
underestimate the number of copies of mRNA for a given gene. However, this
caveat is decreasingly important as $\numi$ increases, and effectively
disappears when $\numi \gg \lambda$ (i.e.\ the number of unique UMIs greatly
exceeds the number of mRNA transcripts for each gene). Assuming a 10 basepair
UMI, this process enables digital quantification of mRNA molecules up to
$4^{10}$ transcripts per gene~\cite{DropSeq}.

Following reverse transcription of the bound mRNA to complementary DNA (cDNA),
we use Polymerase Chain Reaction (PCR) to exponentially amplify the cDNA
library~\cite{sc-rev-tech-bio}. Assume each molecule has some probability
$\pamplify$ of successfully replicating for each round of PCR. Again using the
Poisson marking/thinning property, after $\rounds$ rounds of PCR we have
 \[
   \ntranscripts'_1, \dots, \ntranscripts'_\numi \distiid
   \distPoiss\left(\frac{(1+\pamplify)^\rounds \, \pbead \, \lambda^{(\igene)}}{\numi}\right).
\]

The library is then loaded onto a flow cell for sequencing; we assume that each
transcript hybridizes to the lawn of oligonucleotides with some probability
$\pseq$, yielding
\[
  \ntranscripts''_1, \dots, \ntranscripts''_{\numi} \distiid
  \distPoiss\left(\frac{(1+\pamplify)^\rounds \, \pseq \, \pbead \, \lambda^{(\igene)}}{\numi}\right).
\]

Finally, following sequencing, mRNA counts per gene are quantified by aligning
partial transcripts to a reference genome, with basic error correction to
eliminate singletons and account for sequencing error~\cite{DropSeq, 10X,
sc-rev-tech-bio}, resulting in an overall count for this particular gene,
\[
  x^{(\igene)} = \sum_{i=1}^{\numi} \ind\left[ M''_i > 0 \right].
\]

The distribution of $x^{(\igene)}$ has a closed-form expression:
\[
  x^{(\igene)} &\dist \distBinom\left(
    \numi \; , \;\;
    1-e^{-\dropout \lambda^{(\igene)}} \right)
  \\ \dropout &\defined \frac{(1+\pamplify)^\rounds \, \pseq \, \pbead}{\numi},\label{eq:dropout}
\]
where $\dropout$ is a hyperparameter accounting for gene dropout.
We can model gene-specific dropout by replacing $\dropout$ with
$\dropout^{(\igene)}$ (and decorating the corresponding probabilities in
\eq{dropout} with gene index $\igene$).

Because of experimental dropout and the fact that most genes are turned off at
any given time, the observed expression profile for cell $\icell$, $x_\icell$,
is a sparse vector of digital molecular counts in roughly $\nats^{20,000}$ (for
human cells)~\cite{sc-rev-regev, sc-rev-teichmann, sc-rev-tech-bio}. In
practice, we model the subset of most variable genes for a given dataset, since
low variance genes provide little information to resolve cells along a
trajectory.

\section{P\'olya-gamma augmentation for latent Gaussian states with binomial emissions}\label{app-pg}

We would like to leverage P\'olya-Gamma augmentation
to endow $p(x_\icell \mid \lambda_\icell)$ binomial likelihoods with
(Gaussian) conjugacy.
The P\'olya-Gamma (PG) trick~\cite{pg} applies to the situation
\[
  \lambda \sim \distNorm(\mu, \Sigma) \; ; \;\;\;
  x \mid \lambda \sim \distBinom\left(N , \frac{1}{1 + \exp(-\lambda)} \right).
  \label{eq:pg}
\]
PG augmentation
relies on re-writing the $1 / [1 + \exp(-\lambda)]$ parameterizing the
binomial distribution as $\cosh$, and then recognizing that $\cosh$ is
related to the moment generating function of a PG random
variable~\cite{pg}. If we were to choose a simple positive link function
$\link$ like $\exp(\cdot)$, we would end up with a $\sinh$. Instead:
consider choosing a more suitable link function.

Given the setup
\[
  \lambda_\icell \sim \distNorm(\mu_\icell, \Sigma_\icell) \; ; \;\;\;
  x_\icell \mid \lambda_\icell \sim \distBinom\left( \numi, 1 - \exp\left[
      -\dropout \cdot \link(\lambda_\icell) \right] \right) \label{eq:xbinom}
\]
for a single cell $\icell$'s $\ngenes$-dimensional gene expression profile
$x_\icell$, $\ngenes$-dimensional proto-rates $\lambda_\icell$, and generic mean
and covariance (in the case of $\lambda_\icell$ sampled from a DDT, these would
be given by Gaussian diffusion parameters of the tree),
we can define positive link function $\link$ as
follows:
\[
  \link(\lambda) \defined -\log\left( 1 - \frac{1}{1 + \exp(-\lambda)} \right).
\]
This equation comes from setting the binomial probability we have (\eq{xbinom})
equal to the binomial probability form we need for PG augmentation (\eq{pg}) and
solving for $\link(\cdot)$.
Ignoring dropout parameter $\dropout$ momentarily, observed expression
conditioned on location along the tree can now be rewritten as
\[
  x_\icell \mid \lambda_\icell \sim \distBinom\left( \numi, \frac{1}{1 +
      \exp(-\lambda_\icell)} \right).
\]
Then, with the PG trick, we introduce a $\ngenes$-dimensional auxiliary
variable $\omega_\icell$ per cell, such that we can alternatingly sample
\[
  \omega_\icell \mid \lambda_\icell &\sim \distPG(\numi, \lambda_\icell) \label{eq:pg1}
  \\
  \lambda_\icell \mid x_\icell, \omega_\icell &\sim \distNorm(m_{\omega,
    \icell}, V_{\omega, \icell}) \label{eq:pg2}
\]
for covariance and mean
\[
  V_{\omega, \icell} &= (\Omega_\icell + \Sigma_\icell\inv)\inv
  \\
  m_{\omega, \icell} &= V_{\omega, \icell}(\Sigma_\icell\inv \mu_\icell +
  x_\icell - \frac{\numi}{2})
\]
where $\Omega_\icell \defined \diag(\omega_\icell)$.

Now consider the effect of dropout parameter $\dropout$. Setting
\[
  1 - \exp\left[ -\dropout \, \link(\lambda) \right] \equiv \frac{1}{1 +
    \exp(-\dropout \, \lambda)}
\]
(where the RHS is the desired form for PG augmentation), we obtain
\[
  \link(\lambda) = -\log\left( 1 - \frac{1}{1 + \exp(-\dropout \, \lambda)}
  \right) / \dropout.
\]
In other words, the link function now depends on the dropout parameter
(and the PG equations, \eq{pg1}-\eq{pg2}, apply to $\dropout \, \lambda$
rather than $\lambda$).

Note that this has the effect of scaling a normally-distributed variable
($\lambda$) by a scalar ($\dropout$). Thus, the dropout parameter $\dropout$ can
be absorbed by the diffusion variance $\DDTvar$ (and therefore learned, per gene
or per tree, by placing a conjugate inverse gamma prior over $\DDTvar$).
That is, we directly model the scaled proto-rates ($\dropout \, \lambda$) as
latent values sampled along the tree, rather than explicitly setting $\dropout$
and modeling its effects.

\section{Triplet metric for comparing cell distributions over trees}\label{app-triplet}

We need a metric to compare trees, with the challenge that trees may be of
differing depths, such that node-matching is difficult or impossible. These
desiderata motivate the ``triplet metric,'' which
abstracts away the underlying tree to prioritize 
the topology of the cells themselves and is 
agnostic to tree size.

We define the triplet metric between trees $\tree, \tree'$ over $\ncells$ cells as
\[
\tripletmetric(\tree, \tree') =
\sum_{\{a,b,c\} \in\,\text{cells}}
\indicator\big\{ \text{outlier}_\tree(a,b,c) = \text{outlier}_{\tree'}(a,b,c)
\big\} \bigg/ \binom{\ncells}{3}
\]
where a triplet outlier is defined as the most distant cell of the three (i.e., the set difference between the triplet and the two cells that
share the smallest pairwise path distance).
Here, distance is given by the branch length (in pseudotime) required to traverse the
tree from one cell to another (traveling via the most recent common
ancestor node if not on same branch).
This approach is loosely inspired by the use of summed branch lengths in
phylogenetic literature to denote similarity between leaves (e.g.,\ species)~\cite{phylo}.
Thus, $\tripletmetric$ is a metric from
$0$ (no triplets agree) to $1$ (all triplets agree). In practice, we randomly
subsample cell triplets for efficiency in order to approximate $\tripletmetric$.

\end{document}